\documentclass[%
superscriptaddress,
showpacs,
nofootinbib,
 amsmath,amssymb,
pra,
floatfix,
twocolumn
]{revtex4-1}

\usepackage{graphicx}% Include figure files

\usepackage{bbm}

\usepackage[]{hyperref}

\newcommand{\ket}[1]{| {#1} \rangle \,}

\begin{document}

%\preprint{}

\title{Momentum-dependent pseudo-spin dimers of coherently coupled
  bosons in optical lattices}

\author{Chiara Menotti}%\email{chiara@science.unitn.it}
\affiliation{INO-CNR BEC Center and Dipartimento di Fisica, Universit\`a di Trento, 38123 Povo, Italy}
\author{Fabrizio Minganti} %\email{minganti@clipper.ens.fr}
\affiliation{Ecole Normale Sup\'{e}rieure, International Center of
  Fundamental Physics, Department of Physics, 24 rue Lhomond, F75005
  Paris, France} \affiliation{INO-CNR BEC Center and Dipartimento di
  Fisica, Universit\`a di Trento, 38123 Povo, Italy}
\author{Alessio Recati}%\email{recati@science.unitn.it}
\affiliation{Technische Universit\"at M\"unchen, James-Franck-Stra{\ss}e 1, 85748 Garching, Germany}
\affiliation{INO-CNR BEC Center and Dipartimento di Fisica, Universit\`a di Trento, 38123 Povo, Italy}

\date{\today}

\begin{abstract}

We study the two-body bound and scattering states of two particles in
a one dimensional optical lattice in the presence of a coherent
coupling between two internal atomic levels.  Due to the interplay
between periodic potential, interactions and coherent coupling, the
internal structure of the bound states depends on their center of mass
momentum. This phenomenon corresponds to an effective
momentum-dependent magnetic field for the dimer pseudo-spin,
which could be observed in a chirping of the precession frequency
during Bloch oscillations.  The essence of this effect can be easily
interpreted in terms of an effective bound state Hamiltonian.
Moreover for indistinguishable bosons, the two-body eigenstates can
present simultaneously attractive and repulsive bound-state nature or
even bound and scattering properties.

\end{abstract}


\maketitle

% \tableofcontents

\section{Introduction}

In the recent years it has been demonstrated that ultracold atoms
loaded in optical lattices provide an ideal realization of lattice
Hamiltonians \cite{jaksch2005cold,RevModPhys.80.885}.  The control of the system parameters, in
particular of the ratio between interactions and kinetic energy, has
allowed the experimental achievement of the most famous Mott insulator
to superfluid phase transition in 2002 \cite{Bloch-MottSF}.
From there on, the number of implementations of Hubbard-like models
using cold gases in optical lattices has undergone an incredible
growth (see, e.g., \cite{BookMaciek}): it is now possible to mimic
single- and multi-species Bose and Fermi-Hubbard models, extended
Hubbard models by using dipolar gases \cite{FranceEHM}, and spin chain
models \cite{greiner2011,esslinger2013,blochXXZ}. More recently,
Hubbard models characterised by non-trivial topology, synthetic
magnetic field, artificial gauges
\cite{BlochButterfly,KetterleButterfly}, or synthetic dimensions
\cite{Leosintdim} have been realized.

Most of the theoretical studies on lattice Hamiltonians regard
many-body or single particle properties, especially in the case of non
trivial topology.
On the other hand, lattice Hamiltonians show interesting features also
as far as the few body physics is concerned.  Indeed, when the effect
of interactions is combined with the presence of the lattice,
composite objects on a lattice behave very differently with respect to
their counterpart in free space \cite{MattisFewBody}.
First of all the dispersion relation, i.e. the effective mass of bound
particles depends on the binding energy.  This is due to the
impossibility in the lattice of separating the relative motion and the
center-of-mass degrees of freedom. Moreover the spectrum of the
scattering states, the so-called essential spectrum, is bounded from
below and from above. Therefore bound states can exist both for
attractive and for repulsive interactions \cite{Hubbard1963,
  MattisFewBody}.  The existence of ``exotic" repulsive bound pairs
(RBP) has been directly observed for the first time a few years ago
for an ultra-cold Bose gas in an optical lattice
\cite{winkler2006repulsively}.

\begin{figure}
\includegraphics[width=0.42 \textwidth]{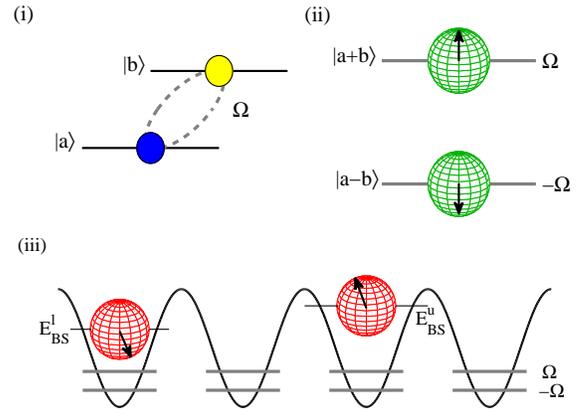}
\caption{Schematic representation of two coherently coupled
  interacting bosons in an optical lattice for $U_a=U_b=U$: (i) two
  internal levels $\ket{a}$ and $\ket{b}$ coherently coupled by a
  resonant coupling $\Omega$; (ii) single-particle eigenstates of
  $\Omega$ at energies $\pm \Omega$ provided by the internal state
  superpositions $\ket{a\pm b}$; in the Bloch sphere representation,
  the two single-particle states $\ket{a\pm b}$ point respectively in
  the positive and negative vertical direction; (iii) repulsive lower
  and upper two-body bound states respectively at energies
  $E_{BS}^{l,u}$: for $U=U_{ab}$ the dimers are formed by both atoms
  in the internal states $\ket{a\pm b}$, namely in the two-body
  internal states $\ket{\pm}$; in the Bloch sphere representation for
  the two-body internal states, the states $\ket{\pm}$ corresponds to
  a pseudo-spin pointing respectively in the positive and negative
  vertical direction; for $U \neq U_{ab}$ the internal wavefunctions
  of the dimers are modified and the corresponding pseudo-spins are
  rotated.}
\label{sketch}
\end{figure}

In this work, we consider two atoms with two coherently coupled
internal levels in a one dimensional (1D) optical lattice. The
two-body properties of this system can be investigated analytically,
allowing for a clear understanding of the interplay between the
exchange term (phase coupling) and the intra- and inter-species
interactions (density coupling). The bound states of the system can be
described in terms of a pseudo-spin wavefunction. Thanks to its
spinorial structure, the bound states can, in certain regimes, show
the co-presence of repulsive and attractive character or the
co-presence of free (delocalized) and bound (localized) nature. Those
features are absent in the single species situation. Moreover, we show
that an {\sl effetive momentum-dependent magnetic field} emerges for
the pseudo-spin of the dressed dimers, by means of a coupling between
the internal state of the bound state and its center-of-mass motion.
The emergent coupling between internal and external degrees of freedom
for the bound states is due to the different dispersion relations of
the bare dimers in the different coupling channels and is very easily
understood in terms of dressed bound states.  The same effect also
emerges in a dimer of two distinguishable particles, provided that at
least one of them possesses two coherently coupled internal levels.

The paper is organised as follows: In Sect.~\ref{model} we describe
the model and make the link to the single species repulsively bound
pairs. We introduce the concept of {\it momentum-dependent} internal
wavefunction for the dimers.  We present an intuitive explanation of
the physics behind this effect using an effective Hamiltonian and an
effective pseudo-spin picture.  In Sect.~\ref{sec:Lippmann-Schwinger},
we solve analytically our model via a Lippmann-Schwinger treatment.
We investigate in detail the properties of the bound states and
discuss the effect of Bloch oscillations on the dimer pseudo-spin
dynamics at the semiclassical level. We describe in detail the case of
hybridized bound and scattering states, which are not present in the
analogous single species model. In Sect.~\ref{asym}, we generalize our
results in the presence of species-dependent hopping and intra-species
interaction and compare the effective model with exact diagonalization
results. In Sect.~\ref{sec:distinguishable}, we show that analogous
results are obtained for dimers of two distinguishable particles, as
long as one of them has two coherently coupled internal levels.
The experimental implementation and relevance of our work in the
cold-gases context is discussed in the conclusions.

\section{Model and Results}
\label{model}

We consider a two-component single-band Bose-Hubbard model with an
exchange term between the two species, characterized by a frequency
$\Omega$.  In second quantization the Hamiltonian reads
\begin{equation}
\hat{H}=\hat{H}_{kin}+\hat{H}_{int}+\hat{H}_{\Omega},
\label{full_H}
\end{equation}
where
\begin{eqnarray}
\label{Ham}
\hat{H}_{kin}&=&-J_a \sum_{\langle i,j \rangle} \hat{a}^{\dagger}_i \hat{a}_{j} -J_b
  \sum_{\langle i,j \rangle} \hat{b}^{\dagger}_i \hat{b}_{j}, \\ 
  \hat{H}_{int}&=&\sum_{i}\left(\frac{U_a}{2}
  \hat{n}_{i}^{a}(\hat{n}_{i}^{a}-1) +\frac{U_b}{2}
  \hat{n}_{i}^{b}(\hat{n}_{i}^{b}-1) + U_{ab} \hat{n}_{i}^{a}
  \hat{n}_{i}^{b}\right), \nonumber \\ 
 \hat{H}_{\Omega}&=&\sum_{i}  \Omega(\hat{a}_i^\dagger\hat{b}_i+ \hat{b}_i^\dagger\hat{a}_i). \nonumber
\end{eqnarray}
The operators $\hat{a}_i$ ($\hat{a}_i^\dagger$) and $\hat{b}_i$
($\hat{b}_i^\dagger$) are the annihilation (creation) operators for a
particle on site $i$ in the internal state $a$ and $b$,
respectively. In this work, we set the lattice constant $d=1$.

A possible realization of model (\ref{full_H}) can be provided by
bosonic atoms in a 1D deep optical lattice, with two hyperfine states
resonantly coupled via an external laser field. At the many-body level
a number of properties of this model in different regimes have been
already studied by means of bosonization techniques or using Density
Matrix Renormalization Group numerical approaches (see, e.g.,
\cite{GiamarchiCC,RecatiCCLattice,McCullochCC}).
In the present work we focus instead on the two-body physics and in
particular on the bound states properties.

Let us first recall that in a single species 1D Bose-Hubbard
Hamiltonian an on-site interaction $U_{\alpha}$ creates a dimer at
energy
\begin{eqnarray}
E^{\alpha}_{K}= {\rm sign}(U_\alpha) \sqrt{U_\alpha^2 +
  16 J^2 \cos^2\left(\frac{K}{2}\right) },
\label{Ebs0}
\end{eqnarray}
where $K$ is the center-of-mass momentum.  Depending on the sign of
the interaction parameter $U_{\alpha}$, the bound state lies above or
below the scattering states, whose energies define the essential
spectrum $\Gamma_{ess}(K)=[-4J \cos(K/2),4J \cos(K/2)]$ (see, e.g.,
\cite{ValienteHM}).  For $U_\alpha>0$, the two-body bound state is
usually referred to as repulsively-bound pair
\cite{winkler2006repulsively}. The intuitive explanation behind
repulsively bound pairs is provided by the fact that the interaction
energy of a bound state cannot be converted into kinetic energy, due
to the limited energy bandwidth.  The dimer has a non-trivial
dispersion relation, which becomes more and more flat for increasing
interactions (see, e.g., \cite{MattisFewBody} and references therein).
For $|U_\alpha| \gg J$, the curvature of the dispersion relation at
$K=0$, proportional to the inverse effective mass, is related to
two-body hopping in second order perturbation theory $J_{eff} \propto
J^2/U_\alpha$.

If we have two species $a$ and $b$ without coherent coupling,
i.e. $\Omega=0$ in Eq. (\ref{Ham}), there is a single bound state at
energy $E^{\alpha}_{K}$ given, for $J_a=J_b=J$, by Eq.~(\ref{Ebs0})
with $U_{\alpha}$ replaced by $U_{aa},U_{bb}$ or $U_{ab}$, depending
on whether the two particles are in the same or in different internal
states.
The bound state wavefunction can be written as $\Psi^{\alpha}_K(r)
\;|\alpha\rangle$, where $\Psi^{\alpha}_K(r)$ is the relative motion
wavefunction and $\ket{\alpha}\in \mathcal{B}_{int}$ describes the
internal composition of the bound state, with
\footnote{We define the state $\ket{1_a,1_b}$ to be the symmetric
  superposition for two particles in species $a$ and $b$. The
  antisymmetric combination, which for bosons would lead to odd
  scattering wavefunctions, does not play any role in the problem
  under consideration in this work.}
\begin{eqnarray}
\mathcal{B}_{int} =\left\lbrace \begin{split} 
\ket{2_a,0_b}, \\
\ket{1_a,1_b}, \\
\ket{0_a,2_b}. \\
\end{split}  \right.
\label{Bint}
\end{eqnarray}
The structure of the eigenfunctions is very simple because the
internal state composition $|\alpha\rangle$ does not depend on
$K$
\footnote{Depending on the initialization of the two particles, the
  two-body system might occupy one or the other bound state, being
  namely in a stationary state, or might be prepared in a
  superposition of the two states which would then oscillate during
  time evolution, even in the absence of the coherent coupling
  $\Omega$.}  
.

Adding a finite $\Omega$ leads to a mixing of the previously described
bound states.  Neglecting for the moment the presence of the
scattering continuum and its influence on the bound states, let us
define for the bound states an effective Hamiltonian whose diagonal
elements are given by the bound state dispersions $E^{\alpha}_K$ in
Eq.~(\ref{Ebs0}).  Assuming that the bound state wavefunctions
$\Psi_K^{\alpha}(r)$ do not depend strongly on $|\alpha\rangle$, the
exchange term provides the off-diagonal matrix elements $\langle
2_a,0_b| {\hat H}_{\Omega} |1_a,1_b\rangle = \langle 1_b,1_b| {\hat
  H}_{\Omega} |0_a,2_b\rangle = \sqrt{2} \Omega$.  Hence for each
value of $K$, one obtains an Hamiltonian of an effective
$\Lambda$-system which reads
\begin{eqnarray}
H_{eff}(K) &=& {\bar{E}_K}+
 \begin{pmatrix}
  \delta_K & \sqrt{2}\Omega &  0 \\
  \sqrt{2}\Omega & \Delta_K &  \sqrt{2}\Omega \\
  0 & \sqrt{2}\Omega &  -\delta_K 
 \end{pmatrix},
\label{effective_hamiltonian}
\end{eqnarray}
with $2{\bar{E}_K}=E^a_K+E^b_K$, $2\delta_K=E^a_K-E^b_K$,
$\Delta_K=E^{ab}_K-{\bar{E}_K}$.

We postpone to Sect.~\ref{asym} the discussion of the general case
$J_a \neq J_b$ and $U_a\neq U_b$ and, for the sake of simplicity, we
consider in most of the paper the $\mathbf{Z}_2$ symmetric situation
where $J_a=J_b=J$ and $U_a=U_b=U$. In that case,
$\bar{E}_K=E^a_K=E^b_K$, $\delta_K=0$, $\Delta_K=E^{ab}_K-{\bar{E}_K}$
and the diagonalization of the effective Hamiltonian leads to two
coupled bright states and a dark state not affected by the coupling
$\Omega$. The dark state, at energy $E_{eff}^0(K)={\bar E_{K}}$, is
the coherent superposition of the two intra-species dimers providing
the Bell state ${\ket 0}=(\ket{2_a,0_b}-\ket{0_a,2_b})/\sqrt{2}$ and
does not depend on the center-of-mass momentum. The bright {\it upper}
($u$) and {\it lower} ($l$) bound states, have energies
\begin{equation}
E_{eff}^{u,l}(K)={\bar E_{K}}+\frac{1}{2} \left(\Delta_K\pm \sqrt{\Delta_K^2 +16 \Omega^2}\right).
\label{eig_eff}
\end{equation}
Their internal wavefunction can be written as 
\begin{equation}
\begin{split}
{\ket {u ; l}}_K\propto&(\ket{2_a,0_b}+\ket{0_a,2_b})+\\ 
&\sqrt{2}\left(\frac{\Delta_K}{2\Omega} \pm
\sqrt{\left(\frac{\Delta_K}{2\Omega}\right)^2+1}\right)\ket{1_a,1_b}.
\end{split}
\label{eff-states}
\end{equation}
In general the internal composition of the bright states does depend
on the center-of-mass momentum.  This is physically due to the fact
that interactions try to preserve polarised bound states, while
coherent coupling prefers to force the atoms in a balanced
superposition of the two internal levels.
From Eq.~(\ref{eff-states}) and the reasoning above, in order for the
momentum-dependence of the bound states to be pronounced, one needs
relevant bare intra- and inter-species energy differences $\Delta_K$
combined with $\Omega,U$ and $U_{ab}$ not much larger than the hopping
parameter $J$.
Conversely, the condition $U=U_{ab}$, i.e. $\Delta_K=0$, defines the
$U(1)$ symmetric case 
\footnote{For $J_a=J_b$ and $U_a=U_b$, the system Hamiltonian is
  characterized by the $\mathbf{Z}_2$ symmetry under the exchange of
  $a$ and $b$, corresponding to a rotation of $\pi$ around the
  $\Omega$-axis; under the condition $U_{ab}=U$, the system acquires a
  further $U(1)$ symmetry and becomes invariant for any rotation
  around $\Omega$.}
, where also the bright bound states are characterized by
the $K$-indipendent internal wavefunctions ${\ket \pm}$, corresponding to
both atoms in the superpositions $\ket{a \pm b}$, eigenstates of
$\hat{H}_\Omega$ (see Eq.~(\ref{Bomega})).
  
The effective Hamiltonian in Eq.~(\ref{effective_hamiltonian}) assumes
bound states well separated from the continuum. For repulsive
(attractive) interactions it can happen that the lower (upper) bound
state energy enters the upper (lower) scattering continuum.
Even in this case, many of the conclusions provided for the upper
(lower) bound state by the effettive model may still remain valid (see
Sect.\ref{sec:bound}). However, as we will show in
Sect.~\ref{sec:bscat}, interesting phenomena can occur due to the
hybridisation between lower (upper) bound states and scattering
states.
In particular, the eigenfunctions present scattering character when
projected on the upper (lower) two-body eigenstate of ${\hat
  H}_\Omega$ and bound state character when projected onto the
orthogonal lower (upper) eigenstate of ${\hat H}_\Omega$.

While we are mainly interested in atoms satisfying Bose statistics,
the previous results apply also for two distinguishable particles.  In
this case it is enough that a Rabi coupling between two different
internal levels is present for one of the particles. The effective
Hamiltonian, as well as the exact results obtained via
Lippmann-Schwinger approach, can be easily extended (see
Sect.\ref{sec:distinguishable}). The advantage is that using two
different atomic species or isotopes the range of available parameters
could be broader, possibly allowing to obtain a larger momentum
dependence of the bound states.

\subsection{$S=1/2$ pseudo-spin dynamics}
\label{pseudo_spin_dyn_su2}

Before discussing the exact solution, let us describe the effect of
Bloch oscillations on the pseudo-spin of the dimers.  As discussed in
Sect.~\ref{sec:Lippmann-Schwinger}, in the $\mathbf{Z}_2$ symmetric
case, the bound state Hamiltonian in
Eq.~(\ref{effective_hamiltonian}) separates in a dark manifold and
in a two-dimensional bright manifold spanned by the states ${\ket
  \pm}$ (see Eq.~(\ref{Bomega})).  Introducing a pseudo-spin notation
for the $\ket{\pm}$ states, the dynamics is driven at each $K$ by
the Hamiltonian
\begin{equation}
H'_{eff}(K)=\bar{E}_K+\frac{\Delta_K}{2}+
2\Omega\sigma_z- \frac{\Delta_K}{2}\sigma_x.
\label{h?}
\end{equation}
The density matrix for a pseudo-spin $S=1/2$ can be written as
$\rho=(1+\mathbf{S}\cdot\sigma)/2$.  From Eq.~(\ref{h?}), the
equations of motion for the pseudo-spin, describing a dimer wavepacket
with center of mass momentum $K$, correspond at a semiclassical level
to the dynamics of a spin in an effective magnetic field
\begin{equation}
{d \over dt}\mathbf{S} =\mathbf{h'}_K\times\mathbf{S},
\end{equation}
with $\mathbf{h'}_K=(-\Delta_K,0,4\Omega)$ .

\begin{figure}
\includegraphics[width=0.45 \textwidth]{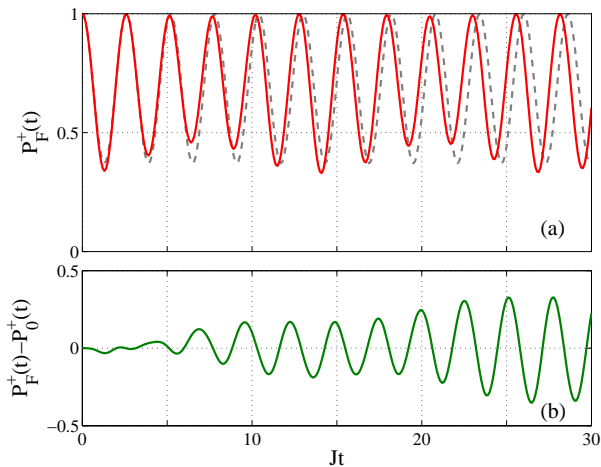}
\caption{Effective pseudo-spin dynamics for a system prepared in the
  two-body state $\ket{+}$ at $t=0$ for $U=J$, $U_{ab}=J/4$ and
  $\Omega=J$: (a) probability $P^+_F(t)$ of finding the system in
  state $\ket{+}$ in the presence of Bloch oscillations driven by a
  constant force $F=J$ (red line) compared to $P^+_{0}(t)$ in the
  absence of driving force (grey dashed line); (b) difference of
  the two probabilities $P^+_{F}(t)-P^+_{0}(t)$  (green line).}
\label{eff_spin_su2}
\end{figure}

In Fig.~\ref{eff_spin_su2}, we plot the probability of finding the
system in the two-body state $\ket{+}$, corresponding to the
expectation value of $S_z$, for two particles prepared at $t=0$ in the
state $\ket{+}$ at $K=0$.  The pseudo-spin of a wavepacket with center
of mass momentum $K$ rotates at a frequency $\omega_L=h'(K)$ and
$\langle S_z \rangle$ oscillates periodically in time. If the center
of mass momentum $K$ varies in time, the consequent spin dynamics
results in a spin precessing in an effective time-dependent magnetic
field $h'(K(t))$.  In particular, Bloch oscillations created by an
external force $F$ induce, in the semiclassical approximation, a
linear variation in time of the center of mass moment
$K(t)=Ft/\hbar$. The effect of the time-dependent $K$ manifests itself
in a chirping of the precession frequency $\omega_L(t)=h'(K(t))$ and
in the appearance of a beating frequency, as reported
Fig.~\ref{eff_spin_su2} (red line in (a) and green line in (b)).

The previous treatment is useful and illustrative in the
$\mathbf{Z}_2$ symmetric case and taking as initial condition a state
belonging to the bright manifold.  In the more general case, e.g.,
starting with a bound state $|2_a,0_b\rangle$, or in the asymmetric
case for which also the dark state is involved in the dynamics, an
effective $S=1$ treatment is required.  Namely one can write directly
the effective Hamiltonian in Eq.~(\ref{effective_hamiltonian})
using the Gell-Mann matrices $\Lambda_i$, $i=1 \dots 8$ as
\begin{equation}
  H_{eff}(K)={\bar{E}_K}+\frac{\Delta_K}{3}+\sum_{i=1}^8
  h_i(K)\Lambda_i,
 \end{equation} 
where we introduced the pseudo-magnetic field
\begin{eqnarray}
&&\mathbf{h}(K) = \\
&&(\sqrt{2}\Omega ,0,(\delta_K-\Delta_K)/2,0,0,\sqrt{2}\Omega ,0,\sqrt{3}(\Delta_K/3+\delta_K)/2). \nonumber
\end{eqnarray}
The density matrix can now be written as
$\rho=(1+\mathbf{S}\cdot\Lambda)/3$.  The equations of motion for the
spin of a wave packet of center of mass momentum $K$ become
\begin{equation}
{d \over dt}S_k =\sum_{l,\;j=1}^8 f^{klj}h_{l}(K)S_j,
\end{equation}
where we use the standard notation for antisymmetric tensor $f^{klj}$
defining the $SU(3)$ Lie algebra
$[\Lambda_l,\Lambda_j]=if^{ljk}\Lambda_k$.

\begin{figure}
%
%\begin{subfigure}{0.33 \textwidth}
%\centering
\includegraphics[width=0.38  \textwidth]{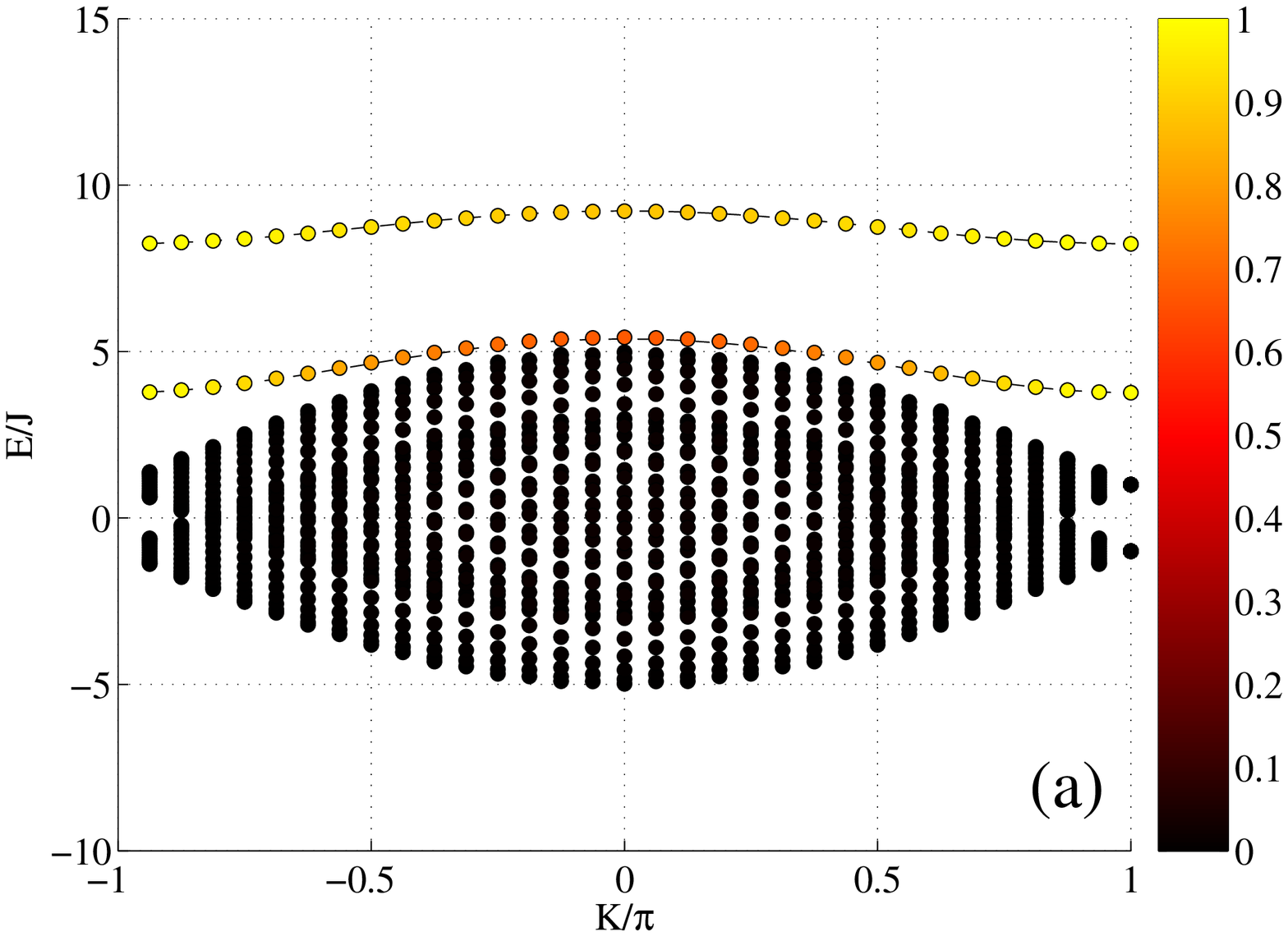}
%\end{subfigure}
%
%\begin{subfigure}{0.33 \textwidth}
%\centering
\includegraphics[width=0.38 \textwidth]{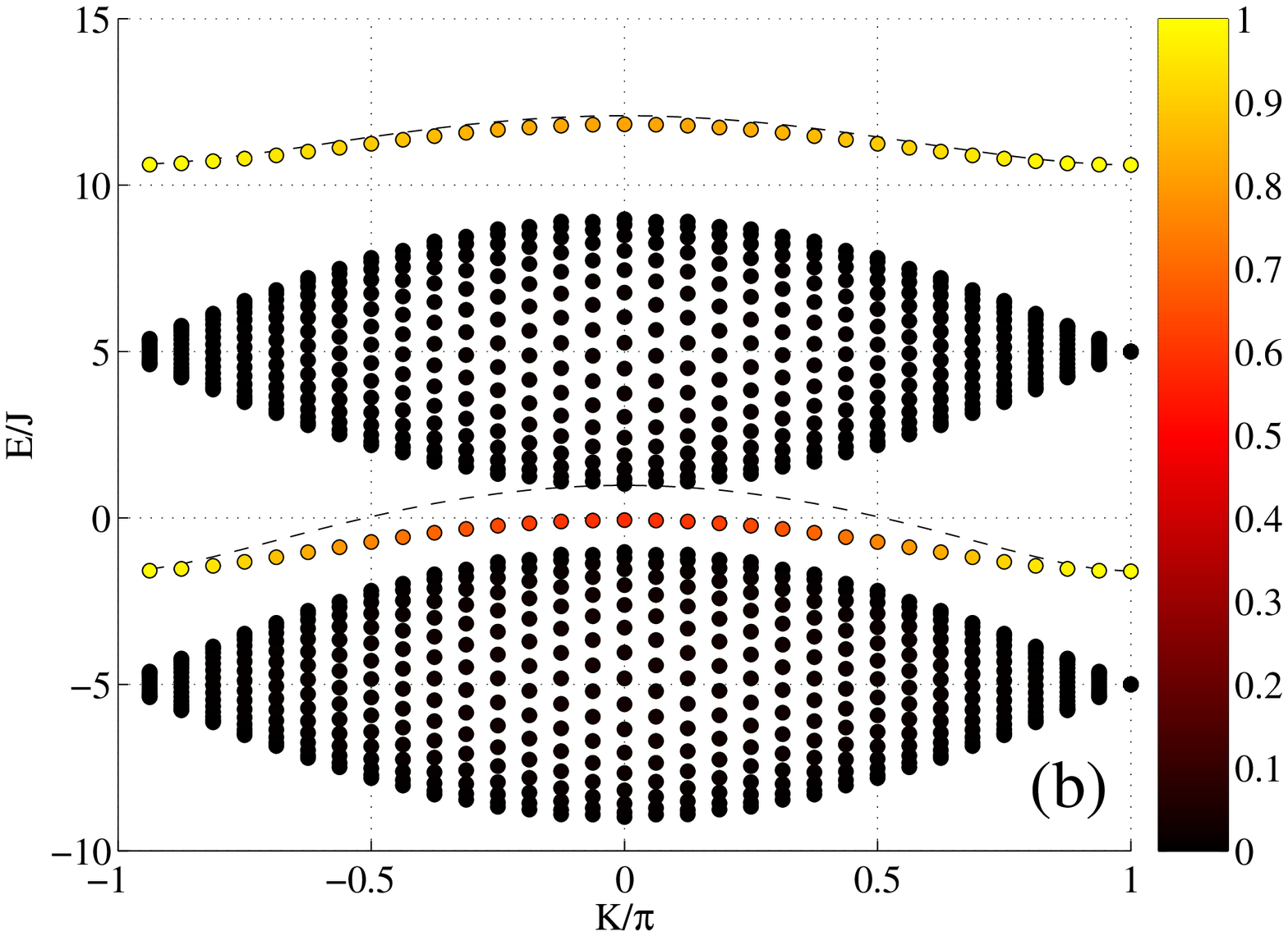}
%\end{subfigure}
%
%\begin{subfigure}{0.33 \textwidth}
%\centering
\includegraphics[width=0.38 \textwidth]{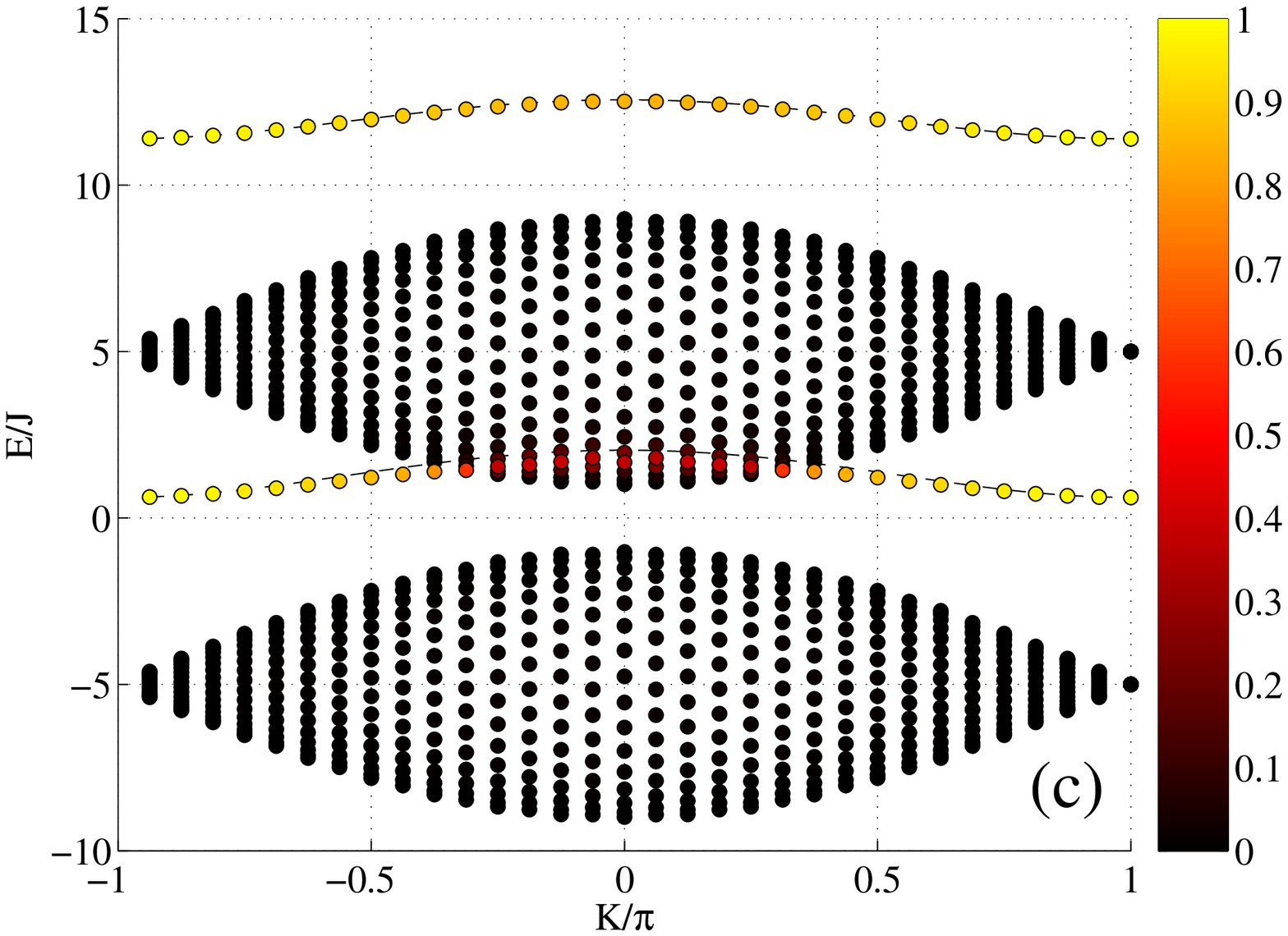}
%\end{subfigure}
%
%\begin{subfigure}{0.33 \textwidth}
%\centering
\includegraphics[width=0.38 \textwidth]{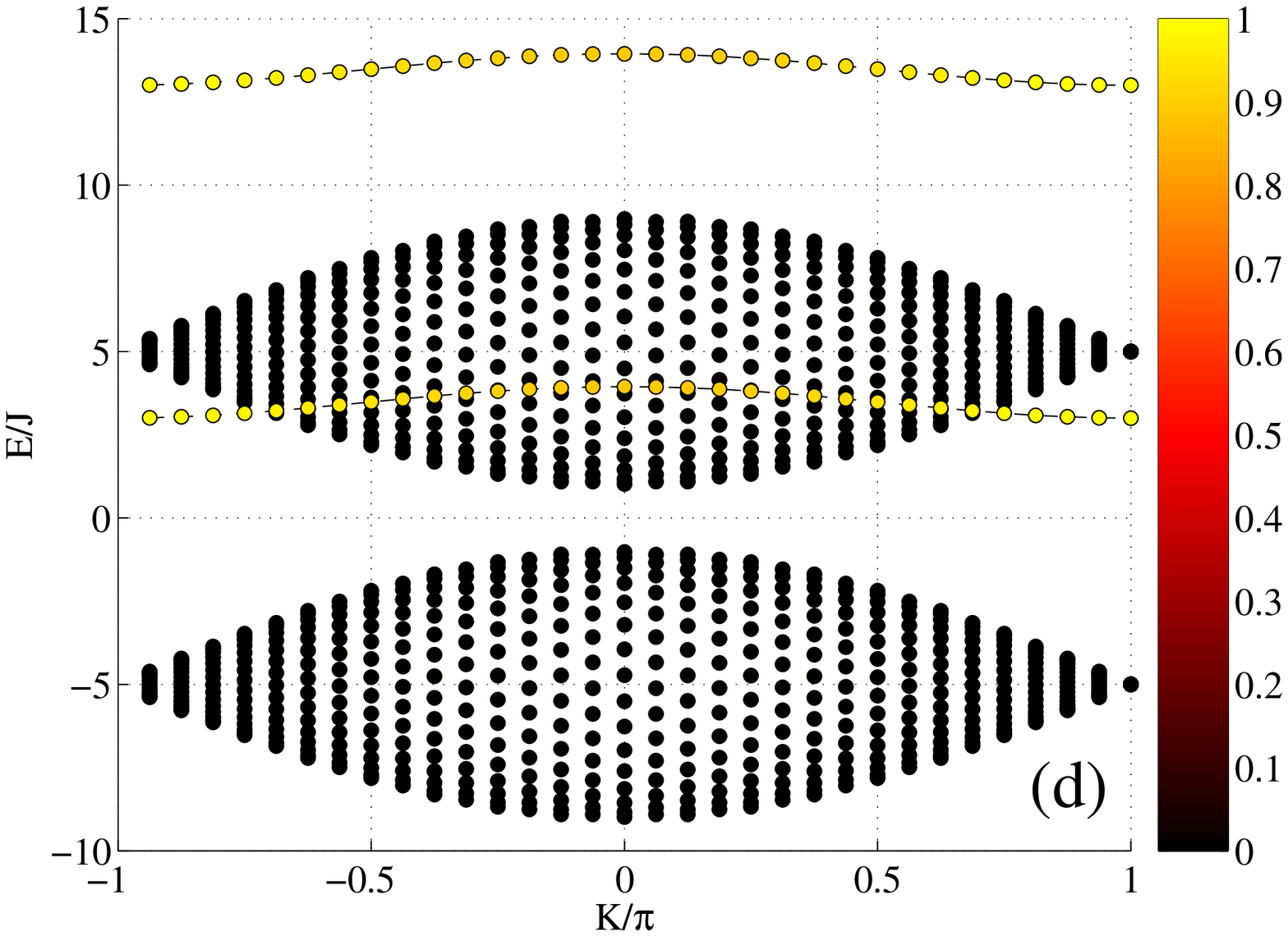}
%\end{subfigure}
%
\caption{Spectrum for $J_a=J_b=J$ and: (a) $U_a=U_b=8J$, $U_{ab}=4J$,
  $\Omega=0.5J$; (b) $U_a=U_b=8J$, $U_{ab}=J$, $\Omega=2.5J$; (c)
  $U_a=U_b=8J$, $U_{ab}=4J$, $\Omega=2.5J$; (d) $U_a=U_b=8J$,
  $U_{ab}=8J$, $\Omega=2.5J$.  For each eigenstate, the population at
  relative distance $r=0$ is plotted in color scale. The dashed lines
  are the bound-state energy predictions of the effective model. For
  clarity, in these figures the dark state spectrum is not shown.}
\label{ex_diag1}
\end{figure}

\section{Lippmann-Schwinger formalism}
\label{sec:Lippmann-Schwinger}

In this section, we solve exactly the two-body problem for Hamiltonian
(\ref{full_H}) using the Lippmann-Schwinger formalism. We restrict
ourselves to the $\mathbf{Z}_2$ symmetric case of $J_a=J_b$ and
$U_a=U_b$, which allows for compact analytical expressions and a clear
interpretation of the results. The more general case is treated in
Sect.~\ref{asym} where the results are obtained by exact
diagonalisation and a generalized effective model.

In order to write the Lippmann-Schwinger equations, it is convenient
to define $\hat{H}_{kin}+\hat{H}_{\Omega}$ to be the free Hamiltonian.
Since we are considering $J_a=J_b$, so that
$[\hat{H}_{kin},\hat{H}_{\Omega}]=0$, the most suited basis is given
by the 2-body internal eigenstates of $\hat{H}_{\Omega}$

\begin{equation}
\mathcal{B}_{\Omega}=\left\lbrace \begin{split} 
\ket{+}& = \frac{\ket{2_a,0_b}+\sqrt{2}\ket{1_a,1_b}+\ket{0_a,2_b}}{2},\\ 
\ket{-}& = \frac{\ket{2_a,0_b}-\sqrt{2}\ket{1_a,1_b}+\ket{0_a,2_b}}{2},\\ 
\ket{0}& = \frac{\ket{2_a,0_b}-\ket{0_a,2_b}}{\sqrt{2}}. \\ 
\end{split}  \right.
\label{Bomega}
\end{equation}
The state $\ket{0}$ corresponds to the dark state discussed above.
Conversely, as it will become clear in Eq.(\ref{forma_U}), the states
$\ket{\pm}$ are coupled to each other by the interaction term.
The essential spectrum $\Gamma^\Omega_{ess} = \cup_\sigma
\Gamma^\sigma_{ess}$ is provided by union of the three scattering
spectra $\Gamma^{\sigma}_{ess}(K)=\Omega_{\sigma}+\Gamma_{ess}(K)$
obtained by shifting the essential spectrum of a single component
Bose-Hubbard model by $\Omega_{\sigma}=2\sigma\Omega$ for
$\sigma=+,\,0,\,-$.

To describe the external degrees of freedom, we follow the standard
procedure of introducing the centre-of-mass coordinate $R=(x_1+x_2)/2$
and the relative coordinate $r=x_1-x_2$ for two particles at lattice
positions $x_1$ and $x_2$.  The center of mass and relative
coordinates do not separate on a lattice, but still the center of mass
momentum $K=k_1+k_2$ is a good quantum number.
Therefore, the eigenstates can be written as spinor wavefunctions
$\sum_\sigma e^{i K R}\psi^{\sigma}_{K}(r)\ket{\sigma}$ with
$\ket{\sigma}\in \mathcal{B}_\Omega$.  Inserting this Ansatz in
Eq.~(\ref{full_H}), we get the discrete Schr\"{o}dinger equation for
the relative motion
\begin{equation}
\begin{split}
\hat{H_r} \psi_{K} ^{\sigma}(r)= 
\left[-2J\tilde{\Delta}_{r}^{K}+\Omega_\sigma \right] \psi_{K}^{\sigma}(r)
+\sum_{\sigma'}U_ {\sigma,\sigma'} \delta_{r,0} \psi_{K}^{\sigma'}(r),
\end{split}
\label{discr_schr_tutto}
\end{equation}
where $\delta_{r,0}$ is the Kronecker-delta and the parametric
dependence of the kinetic energy on $K$ is contained in the discrete
gradient $\tilde{\Delta}_{r}^{(K)} f(r)=\cos(K/2)[f(r+1)+f(r-1)]$. The
interaction matrix reads
\begin{equation}
U_{\sigma,\sigma'}=
\left(
\begin{array}{cccc}
\frac{U+U_{ab}}{2} & \frac{U-U_{ab}}{2} & 0  \\
\frac{U-U_{ab}}{2}& \frac{U+U_{ab}}{2} & 0  \\
0& 0&  U  
\end{array}
 \right) .
 \label{forma_U}
\end{equation}
As already mentioned, interactions mix only the states $\ket{\pm}$
spanning the so-called bright manifold, while the dark state $\ket{0}$
is completely decoupled.

For each value of $K$, let us call
$H_0^{(K)}=-2J\tilde{\Delta}_{r}^{K}+\Omega_\sigma $ the free
Hamiltonian, with eigenstates $|\Phi_K\rangle$ satisfying ${\hat
  H}_0^{(K)} |\Phi_K\rangle = E |\Phi_K\rangle$. The formal solution
of the Lippmann-Schwinger equation reads
\begin{equation}
|\psi_K\rangle=|\Phi_K\rangle+{\hat G}_K(E) {\hat H}_{int} |\psi_K\rangle,
\label{scatt}
\end{equation}
where ${\hat G}_K(E)=(E-\hat{H}_0^{(K)}+i\eta)^{-1}$ is the free
retarded Green's function.

\subsection{Bound states}
\label{sec:bound}

Bound states do exist if the homogeneous equation
$|\psi_K\rangle={\hat G}_K(E) {\hat H}_{int}|\psi_K\rangle$ has
non-zero solutions, or equivalently if there exist values
$E_{BS}\notin \Gamma^\Omega_{ess}$ such that
\begin{equation}
\det[\mathbbm{1}-\hat{G}_K(E_{BS}) \, \hat{H}_{int}]=0.
\label{determinanteenergia}
\end{equation}
In our case the Green's function components read
\begin{equation}
G_K^{\sigma} (r,0,E_{BS})= 
\frac{ {\text{sgn}}(E_{BS}-\Omega_\sigma)(\lambda_K^\sigma)^{|r|}}
{\sqrt{(E_{BS}-\Omega_{\sigma})^2-\beta_{K}^2}},
\label{green_final}
\end{equation}
where we define $\beta_{K}=4J \cos(K/2)$ and
\begin{eqnarray}
\label{lambda}
\lambda_K^\sigma &\equiv&
-\frac{E_{BS}-\Omega_\sigma}{\beta_K} +\\ \nonumber
&& {\text{sgn}}(E_{BS}-\Omega_\sigma)
\sqrt{\left(\frac{E_{BS}- \Omega_\sigma}{\beta_K}\right)^2-1}.
\end{eqnarray}

The dark state gives rise to a bound state equivalent to the one of
the single species Hubbard model with $U_\alpha=U$.  The bright states
are coupled by the interactions and the condition for the existence of
bound states reads 
\begin{equation}
%\det  \Bigg[ 
\Bigg|
\begin{array}{cc}
1- G_{K} ^{+}(0,0,E_{BS})\frac{U+U_{ab}}{2} &  G_{K} ^{+}(0,0,E_{BS}) \frac{U-U_{ab}}{2}\\
&\\
 G_{K} ^{-}(0,0,E_{BS})\frac{U-U_{ab}}{2} & 1 - G_{K} ^{-}(0,0,E_{BS})\frac{U+U_{ab}}{2}  \\
\end{array}
\Bigg|=0.
% \Bigg]= 0.
\label{cond-bs}
\end{equation}
As already noticed in the previous section, for $U=U_{ab}$ the
solution simplifies to two bound states with $K$-independent internal
states $\ket{\pm}$ and effective interaction $(U+U_{ab})/2$.  In the
general case, the two internal states $\ket{\pm}$ are mixed giving
rise to $K$-dependent superpositions. As in the previous section, we
name $u$ ($l$) the upper (lower) bound state.
Typical bound and scattering spectra are reported in
Fig. \ref{ex_diag1}. Since the dark state is completely decoupled, its
spectrum is not shown in the figures.  In panel (a) the parameters are
such that there exists two well defined bound states above the
essential spectrum.  The dashed lines are the approximated bound state
energies given by Eq.~(\ref{eig_eff}), that are almost exact in this
case.  In panel (b) both bound states are well defined for all $K$
values. However due to the vicinity of the lower bound state to the
scattering continuum, the results provided by the effective model
(dashed line) for its energy are not accurate.  In panel (c), the
lowest bound state enters the upper scattering continuum in the
Brillouin zone center. In that case, scattering and bound states get
hybridized, as we will discuss in detail in Sect.~\ref{sec:bscat}. In
spite of the presence of the continuum, which is not accounted for by
the effective model, the prediction for the upper bound state energy
is almost perfect.  For $U=U_{ab}$, one obtains the $U(1)$ symmetric
case, shown in panel (d). In this case, all bound states are perfectly
defined and characterized by the $K$-indipendent $\ket{\pm}$ internal
wavefunctions. In spite of the energy of the lower bound state being
immersed in the upper scattering continuum, there is no
bound-scattering states hybridisation, since the internal wavefunction
of bound and scattering states at the same energy are orthogonal.

\begin{figure}
\includegraphics[width=0.475 \textwidth]{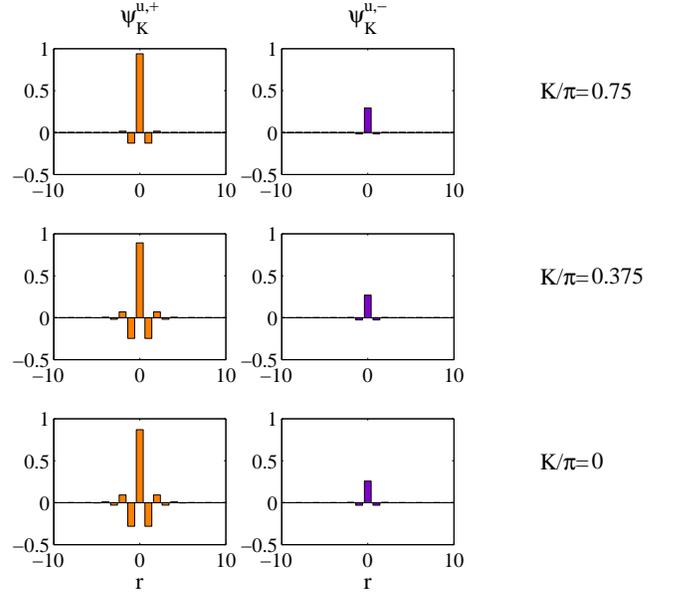}
\caption{Spinorial components in the $\ket{\pm}$ channels of the upper
  bound state for the same parameters of Fig.~\ref{ex_diag1}(b),
  namely $J_a=J_b=J$, $U_a=U_b=8J$, $U_{ab}=J$, $\Omega=2.5J$ and the
  indicated $K$-values.}
\label{upper_wf}
\end{figure}

\begin{figure}
\includegraphics[width=0.475 \textwidth]{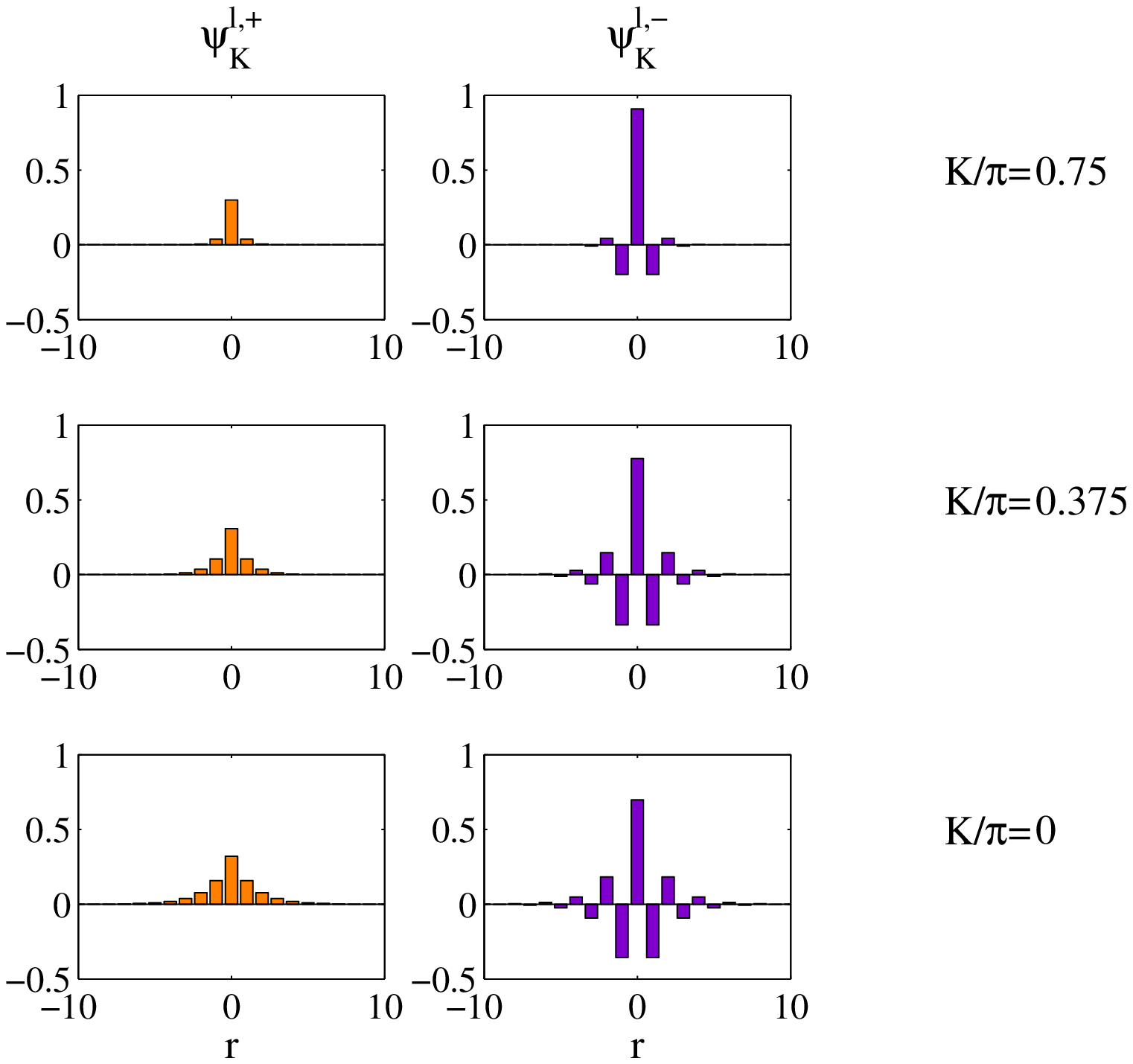}
\caption{Spinorial components in the $\ket{\pm}$ channels of the lower
  bound state for the same parameters of Fig.~\ref{ex_diag1}(b),
  namely $J_a=J_b=J$, $U_a=U_b=8J$, $U_{ab}=J$, $\Omega=2.5J$ and the
  indicated $K$-values.}
\label{lower_wf}
\end{figure}

Once the bound state energies $E^i_{BS}$ are determined by solving
(\ref{cond-bs}), the bound state wavefunctions $\langle{r}
|{\psi^{i}_K}\rangle=\sum_{\sigma=\pm}
\psi^{i,\sigma}_K(r)\ket{\sigma}$, for $i=u,\,l$, can be obtained by
\begin{equation}
\psi_K^{i,\sigma} (r)=G_K^{\sigma}(r,0,E^i_{BS}) \,\sum_{\sigma'} 
U_{\sigma, \sigma'} \psi_K^{i,\sigma'}(0).
\label{scatt}
\end{equation}
The explicit form of the Green's function written in
Eq.~(\ref{green_final}) implies that the amplitudes in the $\ket{\pm}$
states have the following spatial dependence
\begin{eqnarray}
\psi^{\sigma}_K(r) &=& \psi^{\sigma}_K(0) (\lambda_K^\sigma)^{|r|}, 
\label{bs_wf}
\end{eqnarray}
with $\lambda_K^\sigma$ previously defined in Eq.~(\ref{lambda}).
Eventually the ratio at $r=0$ of the spinor components can be written,
e.g., as
\begin{eqnarray}
\label{eq:ratiopm}
\frac{\psi_K^{i,-}(0)}{\psi_K^{i,+}(0)}=
\frac{2 \sqrt{(E^i_{BS}(K)-2\Omega)^2-\beta_{K}^2}}{{\rm sgn}(E^i_{BS}(K)-2\Omega)  (U-U_{ab})}- \frac{U+U_{ab}}{U-U_{ab}}. \;\;\;\;\;\;\;
\end{eqnarray}

From Eq.~(\ref{bs_wf}), it appears clearly that the bound states are
given by two exponentially localized wavefunctions for the $\ket{\pm}$
components with different $K$-dependent decay constants
 \footnote{The condition for the bound state energy not to be in the
   scattering continuum, namely $|E_{BS}-\Omega_\sigma|>\beta_K$,
   ensures real values of the $\lambda$'s.  Moreover, the choice of
   sign in (\ref{lambda}), garantees that $|\lambda_K^\sigma|<1$,
   providing an exponentially decreasing wavefunction. },
as shown in Figs.~\ref{upper_wf} and \ref{lower_wf} for upper and
lower bound state respectively. The sign of $E_{BS}-\Omega_\sigma$
determines the sign of $\lambda_K^\sigma$, which is negative for bound
state energies above the $\sigma$-essential spectrum and positive for
bound state energies below $\Gamma_{ess}^\sigma$. Assuming
$U,U_{ab}>0$, the upper bound state is always above both essential
spectra, implying negative values for both $\lambda_K^\sigma$ and a
$\pi$-paired oscillating wavefunction typical of repulsively bound
pairs \cite{winkler2006repulsively}. Instead the lower bound state can
lie above or between the two essential spectra. In the latter case,
$\lambda_K^-$ is negative, indicating a repulsive character of the
bound state in the $\ket{-}$ internal state, but $\lambda_K^+$ is
positive, indicating that in the $\ket{+}$ internal state the bound
state is actually attractive (see Fig.~\ref{lower_wf}). We remind that
the $\ket{\pm}$ channels describe bound states for two atoms in the
rotated internal states $\ket{a\pm b}$.  Therefore the bound state
wavefunction shows repulsive or attractive nature, or the co-presence
of both, depending on which quantisation direction is chosen in the
measurement of the internal state.

\subsubsection*{$K$-dependent polarization}

In the previous part, we have written the bound state wavefunction in
terms of the ${\ket \pm}$ states and have shown how the two components
depend on the center of mass momentum $K$. The dependence of the
internal wavefunction of the dimers on the center of mass momentum $K$
can be referred to, in a broad sense, as a {\it spin-orbit coupling}
effect.

From the experimental point of view, rather than considering the
${\ket \pm}$ basis, it might be more straighforward to carry out the
measurements using the basis $\mathcal{B}_{int}$, already introduced
to develop the effective model in Sect. \ref{model}. The population in
the different two-body states $\ket{ 2_a,0_b}$, $\ket{1_a,1_b}$ and
$\ket{ 0_a,2_b}$ defines a {\it polarization} vector $\mathcal{P}^i(K)
= [\mathcal{P}^i_{aa}(K); \mathcal{P}^i_{ab}(K);
  \mathcal{P}^i_{bb}(K)]$ for each bound state ($i=u,l$), where
obviously $\sum_\alpha \mathcal{P}^i_\alpha =1$.

\begin{figure}
\includegraphics[width=0.45
  \textwidth]{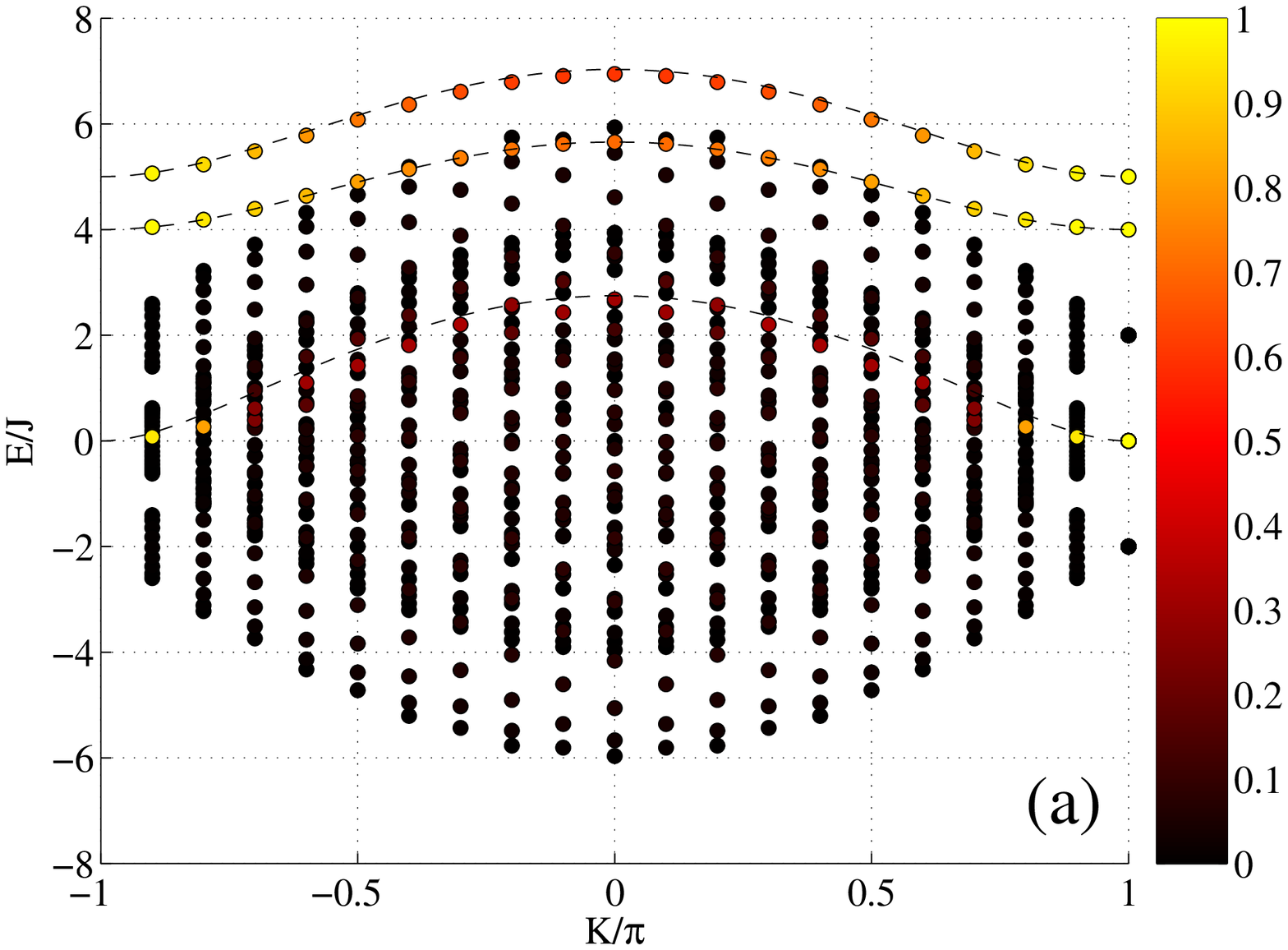} \\
\includegraphics[width=0.415
  \textwidth]{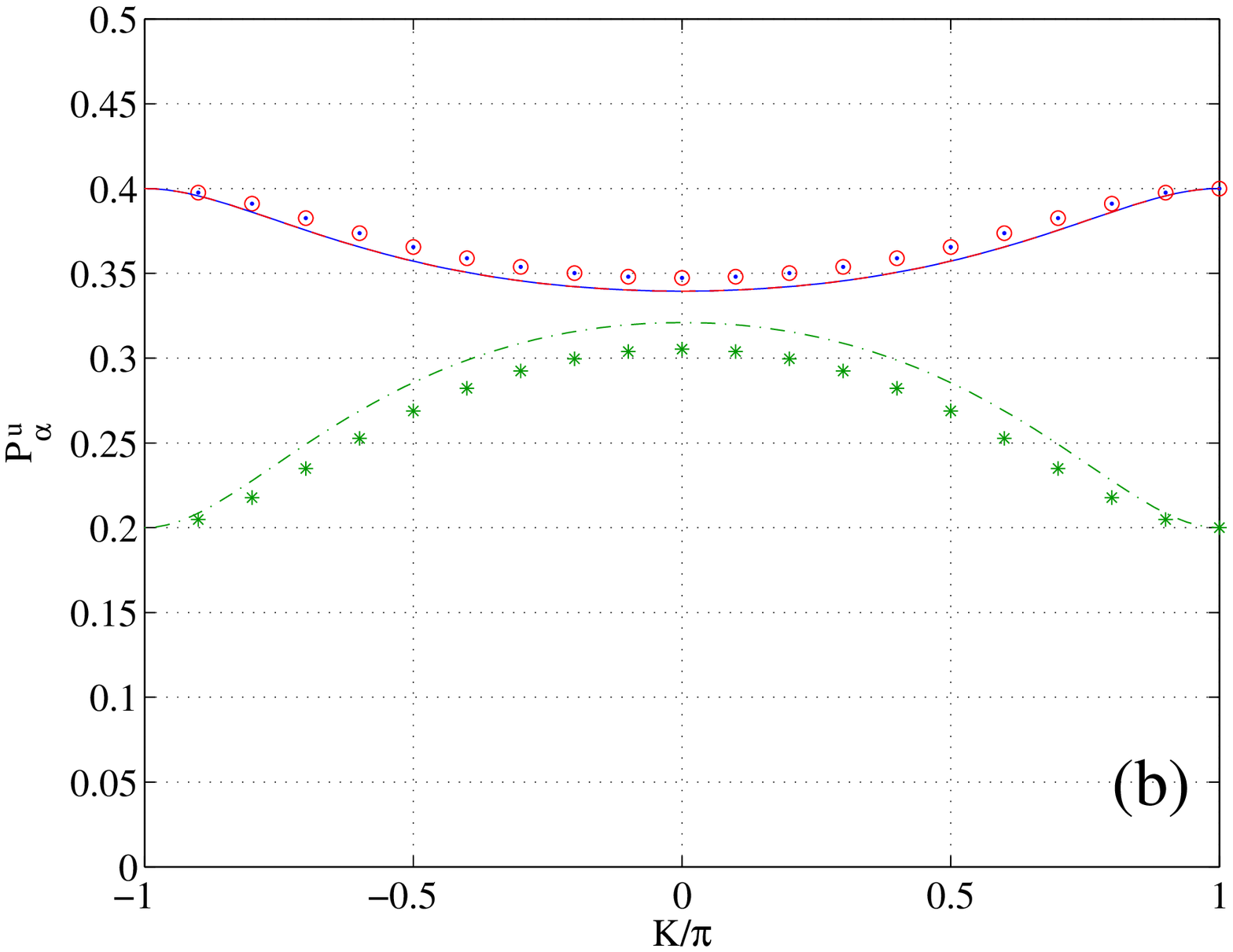} \hspace*{0.9cm}
\caption{(a) Spectrum as a function of $K$ for $J_a=J_b=J$,
  $U_a=U_b=4J$, $U_{ab}=J$, and $\Omega=J$; the dashed lines are the
  results of the effective model, while the dots are the eigenenergies
  of the exact diagonalization; for each eigenstate the population at
  relative distance $r=0$ is plotted in color scale; in this picture,
  the dark state spectrum is included for completeness; (b)
  Polarization ($\mathcal{P}^u_{aa}$ (blue full/dots),
  $\mathcal{P}^u_{ab}$ (green dash-dotted/stars), $\mathcal{P}^u_{bb}$
  (red dashed/circles)) of the upper bound state as a function of
  $K$. The lines are the results of the effective Hamiltonian and the
  symbols the results of exact diagonalization.}
\label{pol_symm}
\end{figure}

To quantify the emergent spin-orbit coupling, we look at the
$K$-dependence of the polarizazion in the upper bound state.  In
Fig.~\ref{pol_symm}, we show a typical situation for $|U-U_{ab}|\sim
\Omega$ and $U$, $U_{ab} \sim J$.  A more detailed analysis, including
also the effect of asymmetries between $J_a,J_b$ and $U_a,U_b$ will be
presented in Sect.~\ref{asym}.

\begin{figure}
\includegraphics[width=0.475 \textwidth]{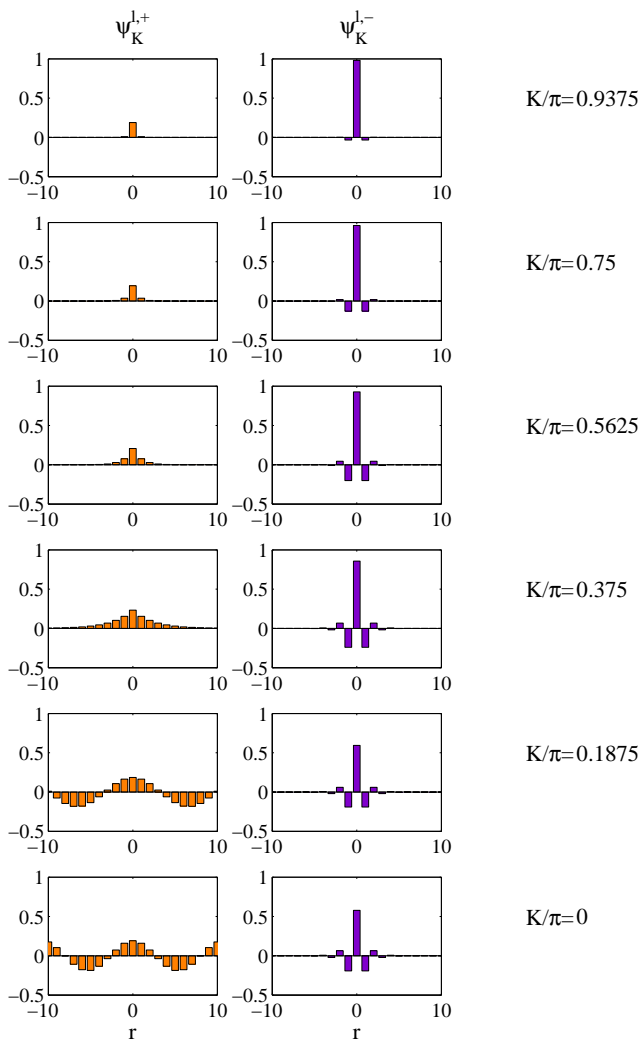}
\caption{Spinorial components in the $\ket{\pm}$ channels of the lower
  bound state for the same parameters of Fig.~\ref{ex_diag1}(c),
  namely $J_a=J_b=J$, $U_a=U_b=8J$, $U_{ab}=4J$, $\Omega=2.5J$ and the
  indicated $K$-values. When the bound state enters the scattering
  continuum, we select for each value of $K$ the most localized state
  in the upper continuum (see colorscale in Fig.~\ref{ex_diag1}(c)).}
\label{pseudo_1_wf}
\end{figure}

\subsection{Bound and scattering states hybridization}
\label{sec:bscat}

As we have already mentioned, for repulsive interaction the lower
bound state can enter the upper scattering spectrum. In other words,
for $U,U_{ab}>0$, there are energies in the scattering spectrum of two
atoms in channel $\ket{+}$ that are immediately close to a pole in the
$T$-matrix for two atoms in channel $\ket{-}$, identified by the
condition $1-G_K^{-}(0,0,E)(U+U_{ab})/2=0$.

In the $U(1)$ symmetric case $U_a=U_b=U_{ab}$, the two internal states
${\ket \pm}$ define two independent channels.  For each channel, at
the energies identified by the poles of the Green's function, one
finds either a scattering or a bound state, like in the single species
case.  Instead, in the general case of different interaction
parameters and in the presence of a coherent coupling between two
species, the two atoms can be found in a superposition of scattering
and bound states. This intriguing concept will be formalized here
below.

The Lippmann-Schwinger equations for $E\in \Gamma^{+}_{ess}$ read
\begin{eqnarray}
\psi_K^+ (r)&=&\cos(k_E r)+G_K^{+,s}(r,0,E) \sum_{\sigma'}\,U_{\sigma, \sigma'} \psi_K^{\sigma'}(0), \\
\psi_K^- (r)&=&G_K^{-}(r,0,E) \sum_{\sigma'}\,U_{\sigma, \sigma'} \psi_K^{\sigma'}(0), 
\end{eqnarray}
where we introduced the internal momentum $k_E$ satisfying
$E=-4J\cos(K/2)\cos(k_E)+\Omega_+$, the Green's function $G_K^{-}$ has
been defined in (\ref{green_final}) and the bare Green's function for
the scattering states reads
\begin{equation}
G_K^{+,s}(r,0,E)=\frac{2}{\beta_K}\frac{\sin(k_E|r|)}{\sin(k_E)}.
\end{equation}

The components of the spinor wavefunction are therefore
\begin{eqnarray}
&&\psi_K^+ (r)=\cos(k_E r)+ \\ &&G_K^{+,s}(r,0,E)\frac{U+U_{ab}}{2}
\left[1+\frac{\frac{(U-U_{ab})^2}{2(U+U_{ab})}G_K^{-}(0,0,E)}{1-G_K^-(0,0,E)\frac{U+U_{ab}}{2}}\right],  \nonumber\\ 
&&\psi_K^- (r)=G_K^{-}(r,0,E) \frac{U-U_{ab}}{2}\frac{1}{1-G_K^{-}(0,0,E)\frac{U+U_{ab}}{2}}. \nonumber
\end{eqnarray}
From the above expressions, it is straightforward to extract some
important properties of the hybridised state: first of all, the
$\psi_K^+$ component presents a delocalized wavefunction, typical of
scattering states; on the other hand, $\psi_K^-$ shows a localized
exponentially decreasing wavefunction typical of bound
states. 

Unless $U=U_{ab}$, the two components feel each other and are
hybrized. The amplitude of the $\psi_K^-$ component is amplified, the
closer the energy is to the bare bound state energy of two atoms in
the two-body state $\ket{-}$. At the same time the $\psi_K^+$
component undergoes a phase shift determined by both by the
interaction coefficient $(U+U_{ab})/2$ in the ${\ket +}$ channel and
the presence of the bound component in the $\ket{-}$
channel. Eventually, in the strongly interacting limit, where the
scattering part strongly dominates over the solution of the
homogeneous equation, the wavefunction is fermionized and developes a
dip at $r=0$. In the presence of the $\psi_K^-$ component, the
strongly interacting limit is reached close to the pseudo-resonance
condition $\sqrt{(E+2\Omega)^2-\beta_K^2} \approx (U+U_{ab})/2$. Those
features are reproduced in Fig.~\ref{pseudo_1_wf} where we plot the
wavefunction varying $K$ from the Brillouin zone edge towards the
center for the parameters corresponding to Fig.~\ref{ex_diag1}(c): At
the Brillouin zone boundary there exists a purely bound state; around
$|K|/\pi \approx 0.3$ the eigenstates become hybridized presenting
both bound and scattering characters.

%%% Effective Hamiltonian %%%
\section{Asymmetric case}
\label{asym}

In this section, we address the most general case where $J_a \neq J_b$
and $U_a \neq U_b$ using exact diagonalization and generalizing the
effective model. As a basis for exact diagonalization calculations, it
is convenient to use all possible distributions in the lattice sites
of two bosons in the single particle atomic states $a$ and $b$
\footnote{Conversely, to perform exact diagonalization in the
  symmetric case, one can reduce the Hilbert space dimension by
  focusing only on the bright subspace spanned by
  $\ket{+}=\ket{2_{a+b},0_{a-b}}$ and
  $\ket{-}=\ket{0_{a+b},2_{a-b}}$. Hence, one can take as a basis all
  possible distributions in the lattice sites of two bosons in the
  internal single particle states $\ket{a \pm b}$.}
.  We label all eigenstates with a well-defined value of the center of
mass momentum $K$ and determine the spectrum for bound and scattering
states. In order to compare the information about the polarization of
the system provided by effective Hamiltonian, we should integrate the
wavefunction provided by the exact solution over relative distance.

The effective Hamiltonian, written in the basis $\mathcal{B}_{int}$,
has the shape shown in Eq.~(\ref{effective_hamiltonian}) presenting
off-diagonal couplings equal to $\sqrt{2} \Omega$, and diagonal
elements, which now explicitely read

\begin{widetext}
\begin{equation}
\left \lbrace
\begin{split}
& E^{a}_K= \mbox{sgn}(U_{a}) \cdot \sqrt{U_{a}^2 + 16 J_a ^2 \cos \left(\frac{K d}{2}\right)^2}, \\
& E^{ab}_K= \mbox{sgn}(U_{ab}) \cdot \sqrt{U_{ab}^2 + 16 \left(\frac{J_a +J_b}{2}\right)^2 \cos \left(\frac{K d}{2}\right)^2+16 \left(\frac{J_a -J_b}{2}\right)^2 \sin \left(\frac{K d}{2}\right)^2 },\\
& E^{b}_K= \mbox{sgn}(U_{b}) \cdot \sqrt{U_{b}^2 + 16 J_b ^2 \cos \left(\frac{K d}{2}\right)^2}.
\end{split} \right.
\label{ebsjajb}
\end{equation}
\end{widetext}
The most relevant difference with respect to the symmetric case is that
the dark and bright manifolds are not defined anymore and in general
three non degenerate bound states are to be expected.

Depending on the specific choice of parameters, the $K$-dependence of
the polarization can be strongly enhanced or strongly suppressed.  To
quantify the amount of spin-orbit coupling present in the system, we
define the visibility ${\mathcal
  V}^u_\alpha=\max_K[\mathcal{P}^u_\alpha]-\min_K[\mathcal{P}^u_\alpha]$
for the different components of the polarization $\mathcal{P}^u$.

\begin{figure}
%
%\begin{subfigure}{0.33 \textwidth}
%\centering
\includegraphics[width=0.4 \textwidth]{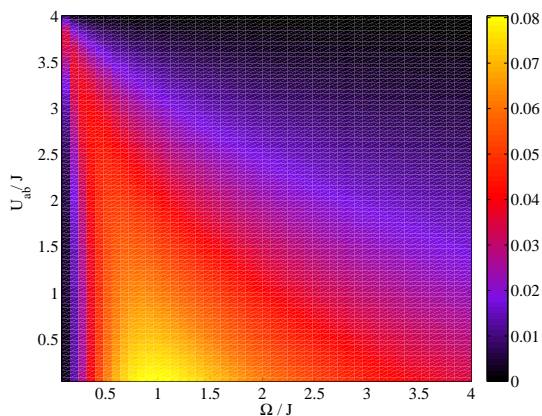}
%\end{subfigure}
%
\caption{Visibility of $\mathcal{P}^u_{aa}$ for $J_a=J_b=J$, and
  $U_a=U_b=4J$ as a function of $\Omega$ and $U_{ab}$. This plot is
  obtained using the effective model, which proves to be very reliable
  in a broad range of parameters.}
\label{vis_pol_symm}
\end{figure}

\begin{figure}
\includegraphics[width=0.4 \textwidth]{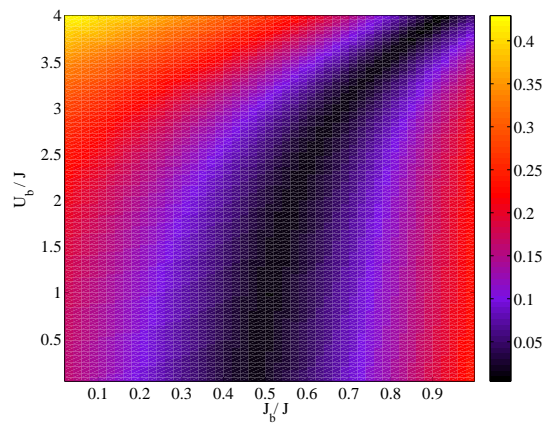}
\caption{Visibility of $\mathcal{P}^u_{aa}$ for $J_a=J$, $U_a=4J$,
  $U_{ab}=J$, and $\Omega=J$ as a function of $J_b$ and $U_{b}$. This
  plot is obtained using the effective model, which proves to be very
  reliable in a broad range of parameters.}
\label{vis_pol_asymm}
\end{figure}

We first consider the symmetric case. From Fig.~\ref{vis_pol_symm}, we
see that for fixed $U_a=U_b=U$, $J_a=J_b$, the visibility of
$\mathcal{P}^u_{aa}$ is largest when $U-U_{ab} \sim 4 \Omega$ and
$U-U_{ab}$ largest. The visibility for the other components follows
straighforwardly from the relations
$\mathcal{P}^u_{bb}=\mathcal{P}^u_{bb}$ and the normalization
condition.  Larger values of $U$ produce very similar behaviours but,
as expected, an overall smaller amount of coupling between internal
and external degrees of freedom.

Taking as a reference the parameters used in Figs.~\ref{pol_symm},
$U_a=4J$, $U_{ab}=J$ and $\Omega=J$, which roughly satisfy the above
conditions, we study how the visibility is affected when asymmetry
between the two species is introduced by varying $J_b$ and $U_b$. The
results for the visibility of $\mathcal{P}^u_{aa}$ is shown in
Fig.~\ref{vis_pol_asymm}.  In this case a non trivial dependence is
observed, in particular as far as the effect of the different
tunneling parameters is concerned.

In general, the effective model produces very reliable results for the
energy and polarization of the upper bound state in a broad range of
parameters, as one can for instance observe in
Fig.~\ref{pol_symm}. This includes the cases where the lowest bound
state is actually immersed in the scattering continuum and
consequently not well described by the effective model.  To get a more
precise idea of the regimes of validity of the effective model, we
calculated the relative difference of the upper bound state energy and
polarization between effective and exact model. At the Brillouin zone
edge $K=\pi$ the effective Hamiltonian always produces exact results,
since there are no scattering states available to hybridize the bound
states.  One obtains exact agreement for $\Omega=0$, which is a
trivial case, and for $U=U_{ab}$ as expected, due to the restored
$U(1)$ symmetry.  The agreement is good for $|U-U_{ab}| \ll \Omega$
and $|U|, |U_{ab}| \ll \Omega$, namely the cases where the dominant
coherent coupling creates bound states in the coherent $\ket{\pm}$
superpositions.  However, it is important to note that even if in the
case of largest spin-orbit coupling, namely $|U-U_{ab}| \sim \Omega$,
the error becomes of the order of several percent, the predictions of
the effective model remain qualitatively reliable.

\section{Distinguishable particles}
\label{sec:distinguishable}

As discussed in the previous sections, the emergent momentum dependent
polarisation is a consequence of the competition between breaking of
Galilean invariace, coherent coupling and interactions. Therefore, a
$K$-dependence of the internal state composition of dressed dimers is
quite a general feature, not unique to the previously discussed
system. In particular, similar concepts can be applied to the case of
two distinguishable particles, as long as at least one of them is
characterised by two internal states, namely $a$ and $b$ coupled by an
exchange term $\Omega$.  Let us called $c$ the state of the second
species or isotope atom.  Interactions are characterized by the two
coefficients $U_{ac}$ and $U_{bc}$.  The different terms of the
2-particle Hubbard-like Hamiltonian in Eq.~(\ref{full_H}) now read
\begin{equation}
\begin{split}
&\hat{H}_{kin}=-J_a \sum_{\langle i,j\rangle} \hat{a}^{\dagger}_i \hat{a}_{j} -J_b
  \sum_{\langle i,j\rangle} \hat{b}^{\dagger}_i \hat{b}_{j} -J_c \sum_{\langle i,j\rangle} \hat{c}^{\dagger}_i \hat{c}_{j}, \\ &
  \hat{H}_{int}=\sum_{i}\left(U_{ac} \hat{n}_{i}^{a}
  \hat{n}_{i}^{c}+ U_{bc} \hat{n}_{i}^{b}
    \hat{n}_{i}^{c}\right), \\ & \hat{H}_{\Omega}=\sum_{i}
  \Omega(\hat{a}_i^\dagger\hat{b}_i+ \hat{b}_i^\dagger\hat{a}_i),
\end{split}
\end{equation}
with $\sum_i({n}_{i}^{a}+{n}_{i}^{b}+{n}_{i}^{c})=2$.  As in
Sect.~\ref{sec:Lippmann-Schwinger}, the convenient basis is provided
by the two-body internal eigenstates of $\hat{H}_{\Omega}$
\begin{equation}
\mathcal{B}_{\Omega}=\left\lbrace
\begin{split}
&\ket{+}=\frac{\ket{1_a,0_b,1_c}+\ket{0_a,1_b,1_c}}{\sqrt{2}},\\
&\ket{-}=\frac{\ket{1_a,0_b,1_c}-\ket{0_a,1_b,1_c}}{\sqrt{2}},
\end{split}
\right.
\end{equation}
with eigenvalues $\Omega_\sigma= \sigma \Omega$.  The eigenstates of
the Hamiltonian can be written as spinor wavefunctions $\sum_\sigma
e^{i K R}\psi^{\sigma}_{K}(r)\ket{\sigma}$ with $\ket{\sigma}\in
\mathcal{B}_\Omega$.

Provided that $J_a=J_b=J_c$, the relative motion for
$\psi^{\sigma}_{K}(r)$ is described again by
Eq.~(\ref{discr_schr_tutto}) and interactions are now given by the
$2\times 2$ matrix
\begin{equation}
U_{\sigma,\sigma'}=
\left(
\begin{array}{cccc}
\frac{U_{ac}+U_{bc}}{2} & \frac{U_{ac}-U_{bc}}{2}   \\
\frac{U_{ac}-U_{bc}}{2}& \frac{U_{ac}+U_{bc}}{2} 
\end{array}
 \right) .
\end{equation}
The analogy with the case described in detail in
Sect.~\ref{sec:Lippmann-Schwinger} is striking and similar conclusions
can be reached.

Analogously, we can write an effective Hamiltonian under the form
\begin{eqnarray}
\hat{H}_{eff}(K)&=&\left(
\begin{array}{cc}
E^{ac}_K & \Omega  \\
\Omega & E^{bc}_K
\end{array}
\right) ,
\end{eqnarray}
where Eq.~(\ref{Ebs0}) with $U_\alpha=U_{ac},U_{bc}$ provides the
diagonal elements.
Two different interaction couplings $U_{ac} \neq U_{bc}$ lead to 
$K$-dependent polarisation for the bound states as previously discussed. 
From the experimental point of view, considering two distinguishable
particles could facilitate the realization of different interaction
and hopping strengths.

\section{Conclusions and perspectives}

In this work we have investigated the bound states of two coherently
coupled interacting bosons.  We have shown an emerging momentum dependent polarisation 
due to the non separability of the centre of mass and
relative coordinates. 
We have derived an effective Hamiltonian, which has allowed us to give
a direct interpretation of the coherently coupled system in terms of
effectively coherently coupled bound states, and tested its validity
against exact solutions based on the Lippmann-Schwinger equation and
exact diagonalization.  We have extended our study to the most general
case where different tunneling parameters and interaction strengths
for the two internal states are considered, in order to possibly
access realistic experimental situations and found the conditions to
optimize the correlations between internal and external degrees of
freedom.
The general character of our results has been stressed by introducing
a similar model for two indistinguishable particles where only one
needs to be dressed by a Rabi coupling.

On the experimental side, many of the ingredients necessary to study
the physics of spin-momentum coupling are already available.  Many
experiments have been already realised for coherently coupled Bose
gas, starting from the seminal investigations in 1999
\cite{Cornell99}, to the Josephson and classical bifurcation
experiments \cite{Zibold2010,OberthalerTwin}, the realisation of
polarisation dependent persistent currents \cite{ZoranSC}, till the
most recent experiments on spin-orbit coupling, artificial gauge
fields and synthetic dimensions, where optical lattices are also
present (see e.g. \cite{DalibardVarenna2014}).  The most difficult
issue is the possibility of realising large enough differences between
intra- and inter-species interactions.  Indeed for the most commonly
used $^{87}$Rb atoms, the difference is very small and Feshbach
resonances are difficult to implement.  A possible idea would be to
consider two different atomic species, as discussed at the end of this
work. Alternatively, one could lift the degeneracy of the hyperfine
states in, e.g. the $F=1$ manifold and use spin-selective microwave
pulses to couple the $F=2$ states in such a way to reduce the
intra-species interaction leaving the inter-species unchanged
\cite{Papoular}.
Another possibility could be to work with spin-dependent lattices as
recently shown in \cite{EsslingerSpinLat}. In this configuration,
aside from breaking $\mathbf{Z}_2$ symmetry, by changing the overlap
between $a$ and $b$ species one could also sensibly tune the
inter-species interaction.

\begin{acknowledgements}
The authors thank J.~Catani, L.~Fallani and W.~Zwerger for useful
discussions and M.~Di Liberto for a careful reading of the manuscript.
This work has been supported by ERC (QGBE grant) and Provincia
Autonoma di Trento.  A.R. acknowledges support from the Alexander von
Humboldt Foundation.
\end{acknowledgements}

\end{document}